\documentclass[pra,twocolumn,floatfix,superscriptaddress]{revtex4-1}
\usepackage{dcolumn,amsmath}
\usepackage{graphicx}
\usepackage{bm}
\usepackage{amssymb}
\usepackage{multirow}

\usepackage{amsfonts}
\usepackage{amsmath}
\usepackage{cancel}
\usepackage{xcolor}
\begin{document}
\title{$\mathcal{P}$,$\mathcal{T}$-odd effects for RaOH molecule in the excited vibrational state}

\author{Anna Zakharova} \email{zakharova.annet@gmail.com} \author {Alexander Petrov}\email{petrov\_an@pnpi.nrcki.ru}
%
\affiliation{St. Petersburg State University, St. Petersburg, 7/9 Universitetskaya nab., 199034, Russia} 
\affiliation{Petersburg Nuclear Physics Institute named by B.P. Konstantinov of National Research Centre
"Kurchatov Institute", Gatchina, 1, mkr. Orlova roshcha, 188300, Russia}
\date{Received: date / Revised version: date}
%
\begin{abstract}{
Triatomic molecule RaOH combines the advantages of laser-coolability and the spectrum with close opposite-parity doublets. This makes it a promising candidate for experimental study of the $\mathcal{P}$,$\mathcal{T}$-violation. Previous studies concentrated on the calculations for different geometries without the averaging over the rovibrational wave function and stressed the possibility that the dependence of the $\mathcal{P}$, $\mathcal{T}$ parameters on the bond angle may significantly alter the observed value. We obtain the rovibrational wave functions of RaOH in the 
ground electronic state and excited vibrational state
using the close-coupled equations derived from the adiabatic Hamiltonian. The potential surface is constructed based on the 
two-component
relativistic CCSD(T) computation employing the generalized relativistic effective core potential (GRECP) for the Radium atom. The averaged values of the parameters $E_{\rm eff}$ and $E_s$ describing the sensitivity of the system to the electron electric dipole moment and the scalar-pseudoscalar nucleon-electron interaction are calculated and the value of $l$-doubling is obtained. 
} 
\end{abstract}
\maketitle
\section{Introduction}
\label{intro}
It is well known from the effects such as mixing and decays of $K$ and $B$-mesons that symmetries under charge conjugation ($\mathcal{C}$), spatial reflection ($\mathcal{P}$) and time reversal ($\mathcal{T}$) are violated \cite{khriplovich2012cp}. All $\mathcal{CP}$ (and thus, $\mathcal{T}$) violation in the Standard model originates from the Cabibbo-Kobayashi-Maskawa (CKM) and Pontecorvo–Maki–Nakagawa–Sakata (PMNS) mixing matrices in the quark and lepton interactions with $W^\pm$-bosons. The other possible source of $\mathcal{CP}$ violation in the strong interaction is severely constrained and that constitutes the so-called strong $\mathcal{CP}$ problem. Moreover, the explanation of the baryon asymmetry in the universe may require new sources of the $\mathcal{CP}$-violation. The popular models of the new physics beyond the Standard Model such as various axion and supersymmetry scenarios predict new $\mathcal{CP}$-violating phenomena. Because of the small coupling constant of the weak interaction and cancellations due to the  Glashow–Iliopoulos–Maiani mechanism, some effects like electron electric dipole moment (eEDM) turn out to be strongly suppressed in the Standard Model compared to the expected new sources $\mathcal{CP}$-violation.

Although the searches of the new physics are popularly associated with collider experiments, the best limits on the eEDM come from high precision atomic and molecular measurements \cite{baron2014order,andreev2018improved}. This allows to put constraints on the new physics on the energies  much higher than accessible on the accelerator experiments. Besides eEDM the same measurements permit study of other $\mathcal{P}$, $\mathcal{T}$ violating phenomena such as scalar-pseudoscalar nucleon-electron interaction \cite{ginges2004violations}, nuclear magnetic quadrupole moment \cite{maison2019theoretical} and interactions with new axionlike particles \cite{maison2020study}.

Some diatomic molecules
with $\Omega=1/2$ ($\Omega$ is projection of total momentum on molecular axis)
that a promising for the $\mathcal{P}$,$\mathcal{T}$-odd interaction measurements, such as RaF \cite{Isaev:2010,GarciaRuiz2020}, YbF \cite{YbFLaserCooled} etc, permit their laser cooling. This allows to increase the coherence time by trapping the molecule and, as result, improve the sensitivity of the experiment. 

In turn diatomic molecules with $\Omega=1$ have closely-spaced $\Omega$-doublet levels.
The energy gap between levels of opposite parity for $\Omega=1$ is much less than for $\Omega=1/2$. Hence polarization of $\Omega=1/2$ requires much stronger electric fields what tends to increase systematic effects.
Also It was shown previously that due to existence of $\Omega$-doublet levels the experiment on ThO \cite{ACME:18,DeMille:2001,Petrov:14,Vutha:2010,Petrov:15,Petrov:17} 
or HfF$^{+}$ \cite{Cornell:2017,Petrov:18} are very robust against a number of systematic effects.

Both the possibility of laser cooling and the existence of the close levels of the opposite parity can be realized with triatomic molecules such as RaOH \cite{Isaev_2017}, YbOH \cite{Kozyryev:17} etc. In this case the role of the $\Omega$-doublets used in the diatomic molecular experiments is overtaken by the $l$-doublets \cite{Kozyryev:17,hutzler2020polyatomic}. Let us elucidate the nature of these levels and their importance for the search of $\mathcal{P}$, $\mathcal{T}$ violation.

The triatomic molecule with linear equilibrium configuration
possesses two bending modes in orthogonal planes.
The superposition of oscillations in these two modes can
be considered as a rotation of a bent molecule characterized
by the rovibrational angular momentum $l$.
Its eigenstates $|+l\rangle$ and $|-l\rangle$ are interchanged
by the parity transformation $\mathcal{P}$. The energy eigenstates
of a free molecule in absence of the external fields are also parity
eigenstates $|\pm\rangle=\frac{1}{\sqrt{2}}(|+l\rangle\pm|-l\rangle)$.
Because the bending modes in the two orthogonal planes are equivalent
the corresponding energy levels are
degenerate. However the Coriolis interaction with stretching modes
results in the split of energies $\Delta E$ of $|\pm\rangle$ states
known as $l$-doubling. Since the dipole moment operator $\hat{d}_z$ is parity-odd,
its expectation values on the parity eigenstates $|\pm\rangle$ vanishes.
The parity symmetry is broken when the molecule is placed into the external
electric field. The perturbed energy eigenstates $|E_\pm\rangle$
become superpositions of the parity eigenstates $|\pm\rangle$ and
the corresponding energies are shifted because of the Stark effect.
For sufficiently high electric fields
$\mathcal{E}\gtrsim \frac{\hbar \Delta E}{d}$ (where $d=\langle l|\hat{d}_z|l\rangle$)
the Stark effect changes from the quadratic regime to the linear one.
At this point the energy eigenstates become to a good degree rovibrational
momentum eigenstates $|\pm l\rangle$ and the molecule is fully polarized
i.e. the dipole moment expectation value reaches maximum. In absence of the $\mathcal{P}$,$\mathcal{T}$-violation, energy levels will not depend on the sign of the total angular momentum projection $M$ on the electrical field axis. Thus the difference between energy shifts for $M=+1$  and $M=-1$ can be used to measure $\mathcal{P}$ and $\mathcal{T}$ violation.
The maximum splitting is given by
\begin{equation}
2E_{\rm eff}  d_e + 2E_{s} k_s,
\label{split}
\end{equation}
where $d_e$ is the value of eEDM, $k_s$ is a characteristic dimensionless constant for scalar-pseudoscalar nucleon-electron interaction. To extract $d_e$ and $k_s$ from the measured
splitting one need to know $E_{\rm eff}$ and $E_{s}$ which are subject of molecular calculations \cite{denis2019enhancement,prasannaa2019enhanced,gaul2020ab}. 

Previous studies of triatomic molecules include calculation of $E_{\rm eff}$ and $E_{s}$
for different geometry of the molecules. However, the final number should be given by averaging over rovibrational wave function what was not made previously. In \cite{prasannaa2019enhanced} the strong dependence of the $E_{\rm eff}$ on the bond angle for YbOH molecule was noted that may significantly affect the observed value. The quantum number $l$ can take values $l= v, v-2,\dots, 1(0)$, where $v$ is quantum number for bending mode. 
Thus the lowest vibrational state with $l-$doubling structure is $v=1$ level, which is of primary interest.

As the necessary
electric field strength is proportional to the value of $l$-doubling, this parameter is important to estimate the applicability of molecule.

The aim of whis work is to calculate $E_{\rm eff}$ and $E_{s}$ for the lowest vibrational levels and l-doubling for $v=1$.

\section{Averaging over nuclear wavefunction}

\begin{figure}[h]
\centering
  \includegraphics[width=0.25\textwidth]{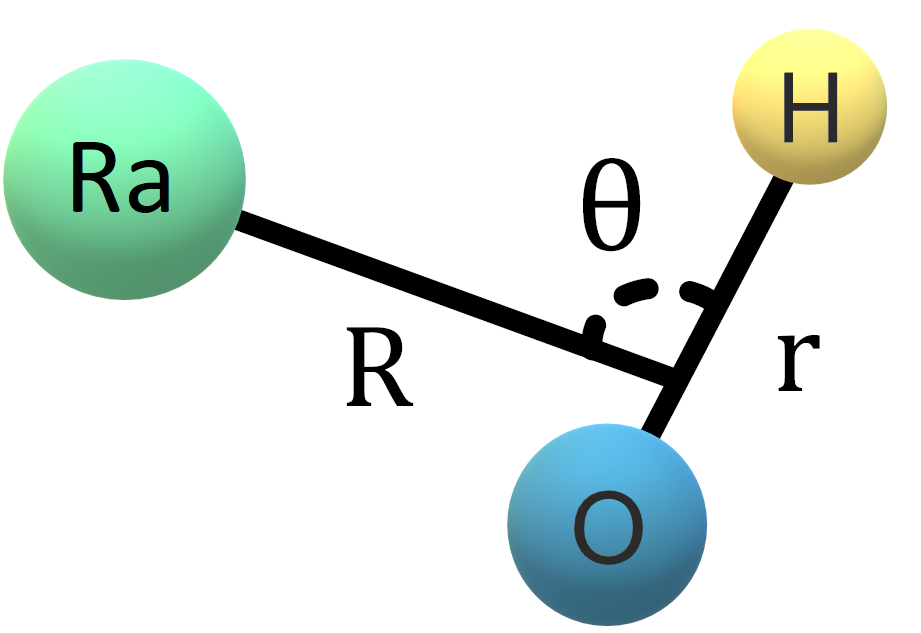}
  \caption{Jacobi coordinates}
  \label{Jacob}
\end{figure}

In the Born-Oppenheimer approximation the total wavefunction of the molecule in Jacobi coordinates (Fig. \ref{Jacob}) takes the form,
\begin{equation}
\Psi_{total}=\Psi_{nuc}(R, \hat{R}, \hat{r})\psi_{elec}(R,\theta|q),
\end{equation}
q means  coordinates of the electronic subsystem,
$\hat{r}$ and $\hat{R}$ are directions of OH axis and Ra - center mass of OH axis respectively, $\theta$ is the angle between above axes, $R$ is Ra - center mass of OH separation. In the current approximation we fix OH ligand stretch at the equilibrium distance $r=1.832 a_0$. This is a reasonable approximation since frequency of OH vibrational mode is about one order of magnitude larger than other vibrational frequencies in RaOH. 

The electronic wavefunction $\psi(R,\theta|q)$ is the solution of the multi-electron Dirac equation in the field of the stationary classical nuclei.
The Hamiltonian for the nuclear motion in the Jacobi coordinates with fixed OH ligand stretch reads as
\begin{equation}
\hat{H}_{nuc}=-\frac{1}{2\mu}\frac{\partial^2}{\partial R^2}+\frac{\hat{L}^2}{2\mu R^2}+\frac{\hat{j}^2}{2\mu_{OH}r^2}+V(R,\theta),
\end{equation}
where $\mu$ is the reduced mass of the $Ra-OH$ system, $\mu_{OH}$ is the reduced mass of the OH ligand, $\hat{L}$  is the angular momentum of the rotation of Radium atom and OH
around their center of mass, $\hat{j}$ is the angular momentum of the rotation of the OH, and $V(R,\theta)$ is the effective adiabatic potential.
The nuclear wavefunction $\Psi_{nuc}(R, \hat{R}, \hat{r})$ is the solution of the Schr\"{o}dinger equation
\begin{equation}
\hat{H}_{nuc}\Psi_{nuc}(R, \hat{R}, \hat{r}) = E \Psi_{nuc}(R, \hat{R}, \hat{r}).
\label{Shreq}
\end{equation}
To solve eq. (\ref{Shreq}) we use expansion
\begin{equation}
\Psi_{nuc}(R, \hat{R}, \hat{r}) = \sum_{L=0}^{L_{max}}\sum_{j=0}^{j_{max}} F_{JjL}(R)\Phi_{JjLM}(\hat{R},\hat{r}),
\label{psiexp}
\end{equation}
where
\begin{equation}
\Phi_{JjLM}(\hat{R},\hat{r}) = \sum_{m_L,m_j} C^{JM}_{Lm_L,jm_j} Y_{Lm_L}(\hat{R})Y_{jm_j}(\hat{r})
\end{equation}
is coupled to conserved total angular momentum $J$ basis set, $Y_{Lm_L}$ is spherical function.
Due to parity conservation the sum $L+j$ must be even or odd for
positive or negative parity respectively.

 The potential surface is expanded in terms of the Legendre polynomials,
\begin{equation}
V(R,\theta)=\sum_{k=0}^{k_{max}}V_{k}(R) P_{k} (\cos\theta)
\label{LegendreExpansion}
\end{equation}
Substituting  wavefunction (\ref{psiexp}) to eq. (\ref{Shreq}) one gets
the system of close-coupled equations for $F_{JjL}(R)$ \cite{mcguire1974quantum}. We found that solution is completely converged for $L_{max}=j_{max} =20$ and $k_{max}=40$

The eEDM and scalar-pseudoscalar nucleon-electron interaction on the molecule can be described by $\mathcal{P}$, $\mathcal{T}$-odd Hamiltonian
\begin{align}
&\hat{H}_{\cancel{\mathcal{PT}}}=\hat{H}_d+\hat{H}_s,\\
&\hat{H_d}=  2d_e\sum_{i}
  \left(\begin{array}{cc}
  0 & 0 \\
  0 & \bf{\sigma_i E_i} \\
  \end{array}\right)\ 
 \label{Hd},\\
&\hat{H_s}=ik_s\frac{G_F}{\sqrt2}\sum_{j=1}^{N_{elec}}\sum_{I=1}^{N_{nuc}}{\rho_I\left(\vec{r_j}\right)Z_I}\gamma^0\gamma^5
\end{align}
where 
$G_F$ is Fermi constant, and $\rho_I$ is the charge density of the $I$-th nucleon normalized to unity,  $\bf{E_i}$ is the inner molecular electric field acting on ith electron, $\bf{\sigma}$ are the Pauli matrices.

Considering these interactions as a small perturbation their impact on the spectrum can be described by the expectation
values of 
\begin{align}
E_{\rm eff}(R,\theta)=\frac{\langle\psi_{elec}(R,\theta)| \hat{H}_d|\psi_{elec}(R,\theta)\rangle}{d_e{\rm sign}(\Omega)},\\ E_s(R,\theta)=\frac{\langle\psi_{elec}(R,\theta)| \hat{H}_s|\psi_{elec}(R,\theta)\rangle}{k_s{\rm sign}(\Omega)},
\end{align}
on nuclear wavefunction (\ref{psiexp}):
\begin{align}
\label{Eeffaver}
E_{\rm eff}=\int dR d\hat{R} d\hat{r} |\Psi_{nuc}(R, \hat{R}, \hat{r})|^2 E_{\rm eff}(R,\theta),\\
\label{Esaver}
E_s=\int dR d\hat{R} d\hat{r} |\Psi_{nuc}(R,\hat{R}, \hat{r})|^2 E_s(R,\theta),
\end{align}

\section{Methods}

We used the Dirac 19 software package \cite{DIRAC19} to calculate molecular orbitals using the Dirac-Hartree-Fock self-consistent field (SCF) method, as well as to construct the potential surface in the coupled-cluster approximation with single, double and perturbative triple excitations (CCSD(T)). The cc-pVTZ basis was employed for O and H atoms. To describe the electronic structure of Ra atom we used a 10-valence electron basis with a generalized relativistic effective core potential (GRECP) \cite{titov1999generalized,mosyagin2010shape,mosyagin2016generalized} including spin-orbit interaction blocks that was developed by the PNPI Quantum Chemistry Laboratory \cite{QCPNPI:Basis}.

The computations were performed for the molecular configurations corresponding to a grid of Jacobi coordinates. The values of $R$ were chosen to cover the span from $3.6\, a.u.$ to $6.0\,a.u.$ with step $0.2\, a.u.$ The values of $\theta$ correspond to the zeroes of Legendre polynomial $P_5$ and two angles for linear configurations i.e. $0^\circ$, $25^\circ$, $57^\circ$, $90^\circ$, $122^\circ$, $155^\circ$ and $180^\circ$. 
\begin{figure}[h]
\centering
  \includegraphics[width=0.45\textwidth]{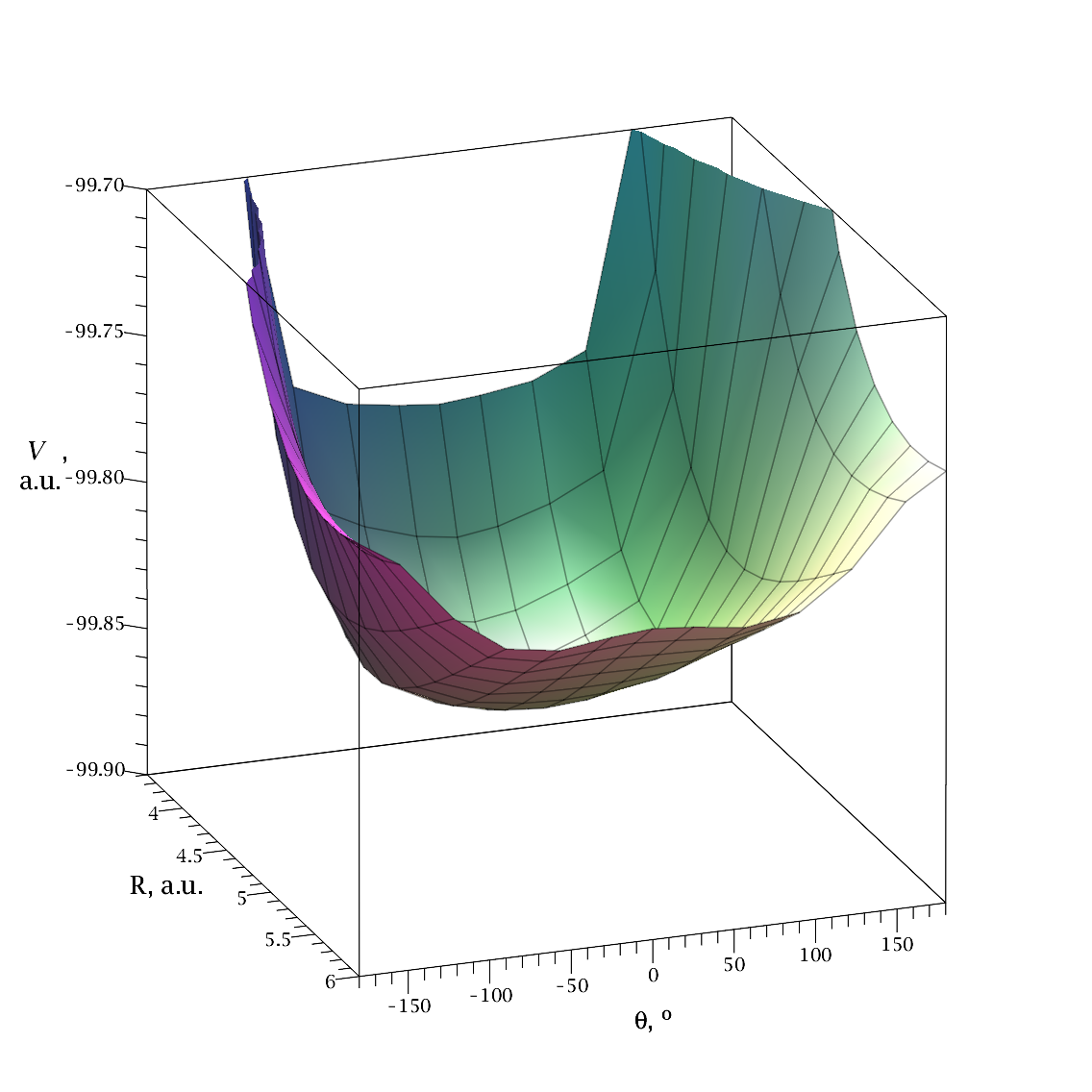}
  \caption{RaOH potential surface $V(R,\theta)$ at CCSD(T) level. The coordinates are introduced in Fig. \ref{Jacob}.}
  \label{fgr:Fig.4}
\end{figure}

Spinors calculated using GRECP have incorrect behavior in the core region. To restore the correct spinor functions, the method of one-center restoration based on equivalent bases, implemented in the MOLGEP program, was applied \cite{Petrov:02,titov2006d,skripnikov2015theoretical}. MOLGEP is restricted to the real two-component molecular orbitals. For this paper we developed the code that applies MOLGEP to compute the matrix elements on the complex orbitals in the Dirac quaternionic representation.

The orbitals obtained in Dirac were used to calculate the matrix elements of properties in MOLGEP. Convolution of the matrix elements computed on the molecular orbitals $\psi_i$ with the one-electron density matrix $\rho^{(1)}_{ij}$ gives the average property value for the electron configuration,
\begin{equation}
\langle \mathcal{O}\rangle=\frac{1}{N_{elec}}\sum_{i,j=1}^{N_{orb}}\rho^{(1)}_{ij}\langle \psi_i|\hat{\mathcal{O}}|\psi_j\rangle.
\end{equation}
The SCF density matrix in the molecular orbital basis was constructed based on the occupation: $\rho^{(1)}_{ij}=\delta_{ij}$ if both indices $i,j$ correspond to the occupied orbitals, and $\rho^{(1)}_{ij}=0$ if any index correspond to the virtual orbital. The correlated single-electron density matrices were obtained for RaOH linear configurations using the CCSD method implemented in the MRCC software package \cite{MRCC2020}. This method was used to calculate the SCF values and the correlation corrections for the P, T-odd parameters $E_{\rm eff}$ and $E_s$. Due to the restrictions of the Dirac-MRCC interface the correlation corrections were computed only for the linear configurations. Because of the small magnitude of these corrections we neglect their dependence on the angle and apply them to the nonlinear configurations.

\begin{figure}[h]
\centering
  \includegraphics[width=0.40\textwidth]{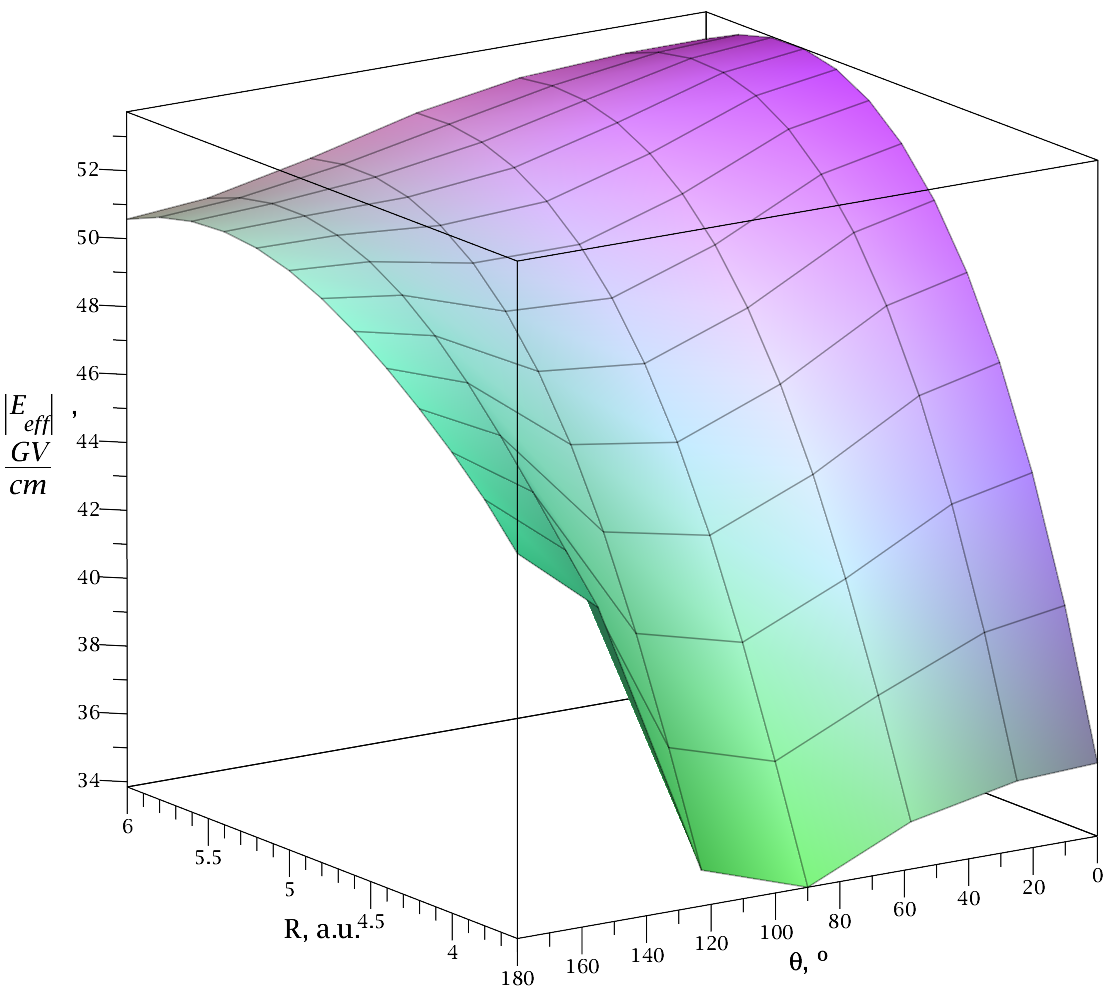}
  \caption{Results for $E_{\rm eff}(R,\theta)$ as a function of Jacobi coordinates at CCSD level}
  \label{fgrWd}
\end{figure}

\begin{figure}[h]
\centering
  \includegraphics[width=0.40\textwidth]{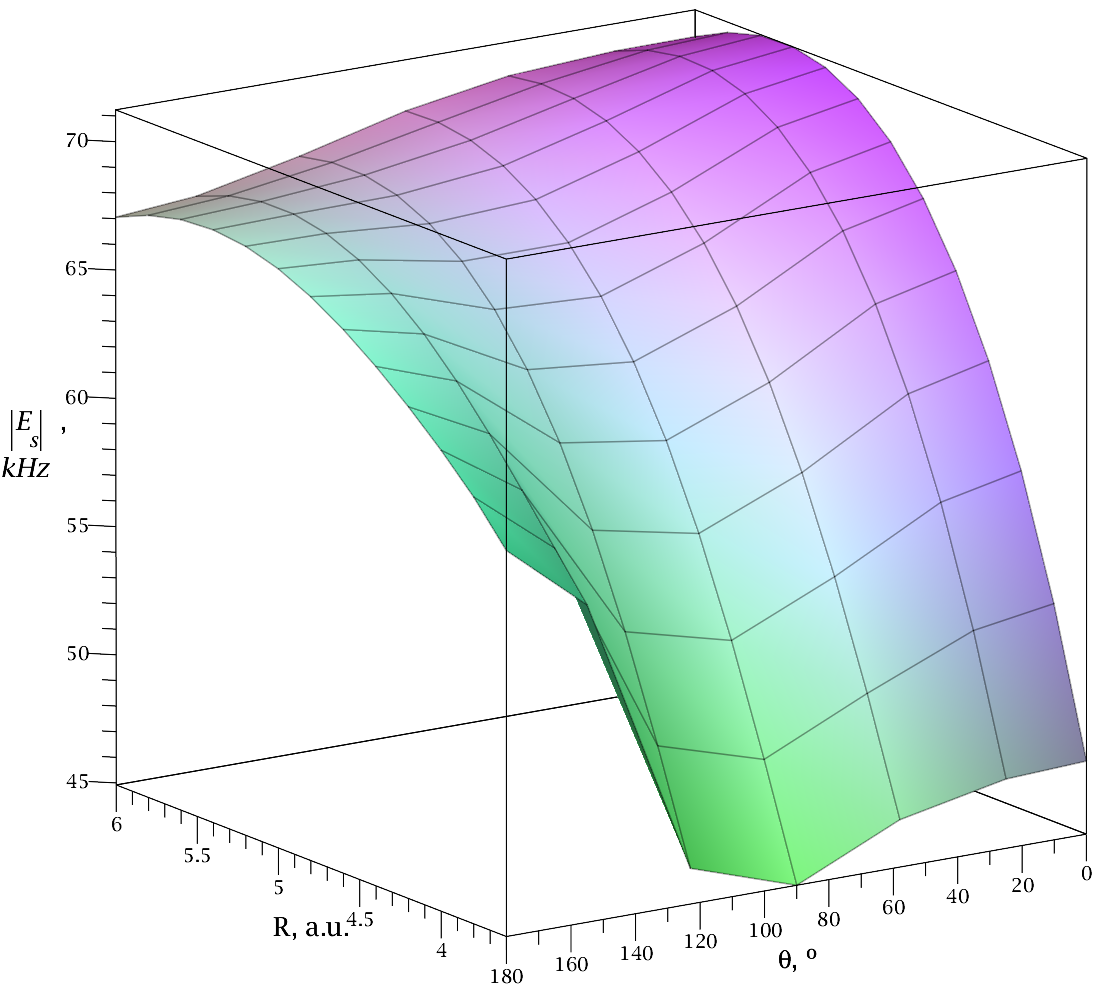}
  \caption{Results for $E_s(R,\theta)$ as a function of Jacobi coordinates at CCSD level}
  \label{fgrWs}
\end{figure}

\begin{figure}[h]
\centering
  \includegraphics[width=0.23\textwidth]{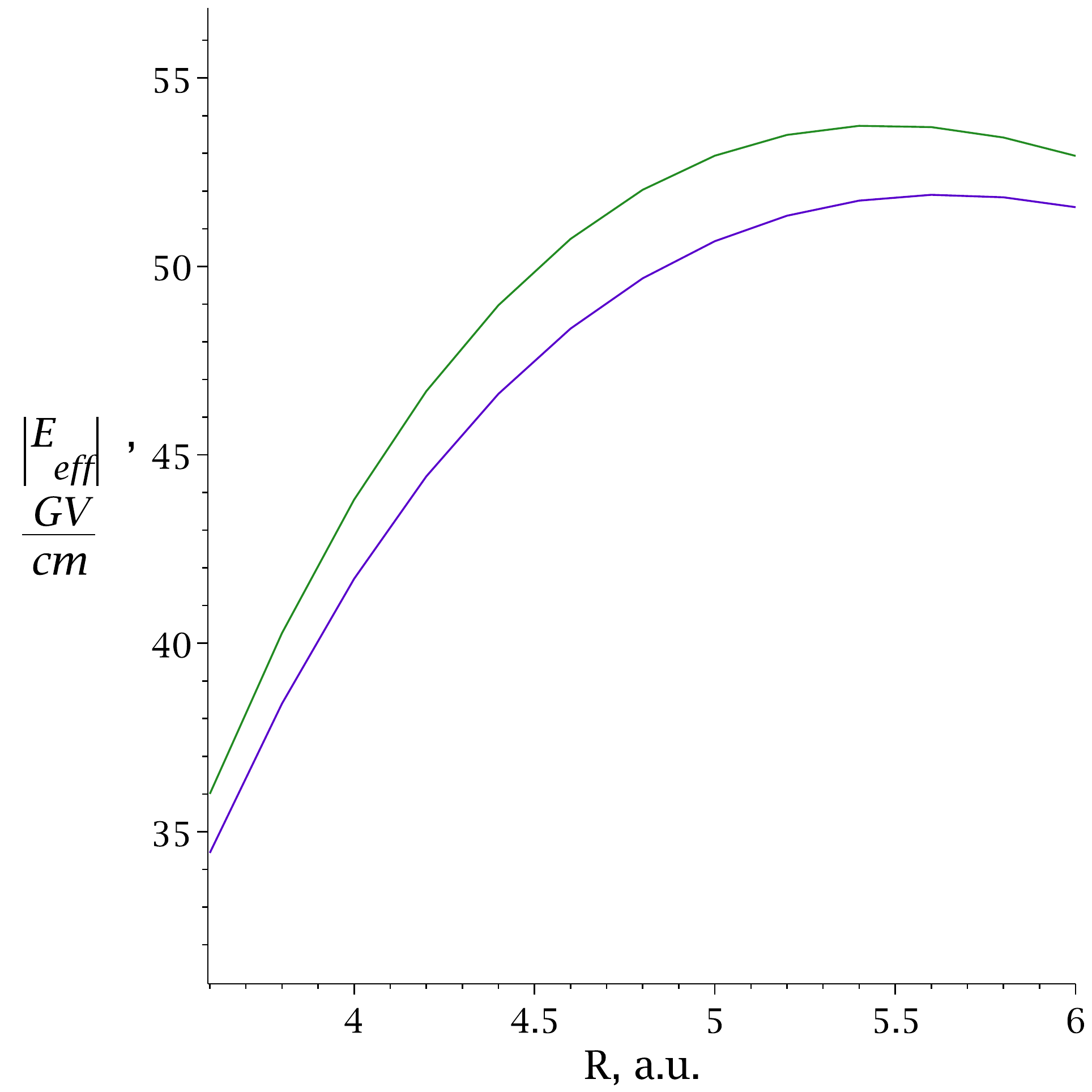}
  \includegraphics[width=0.23\textwidth]{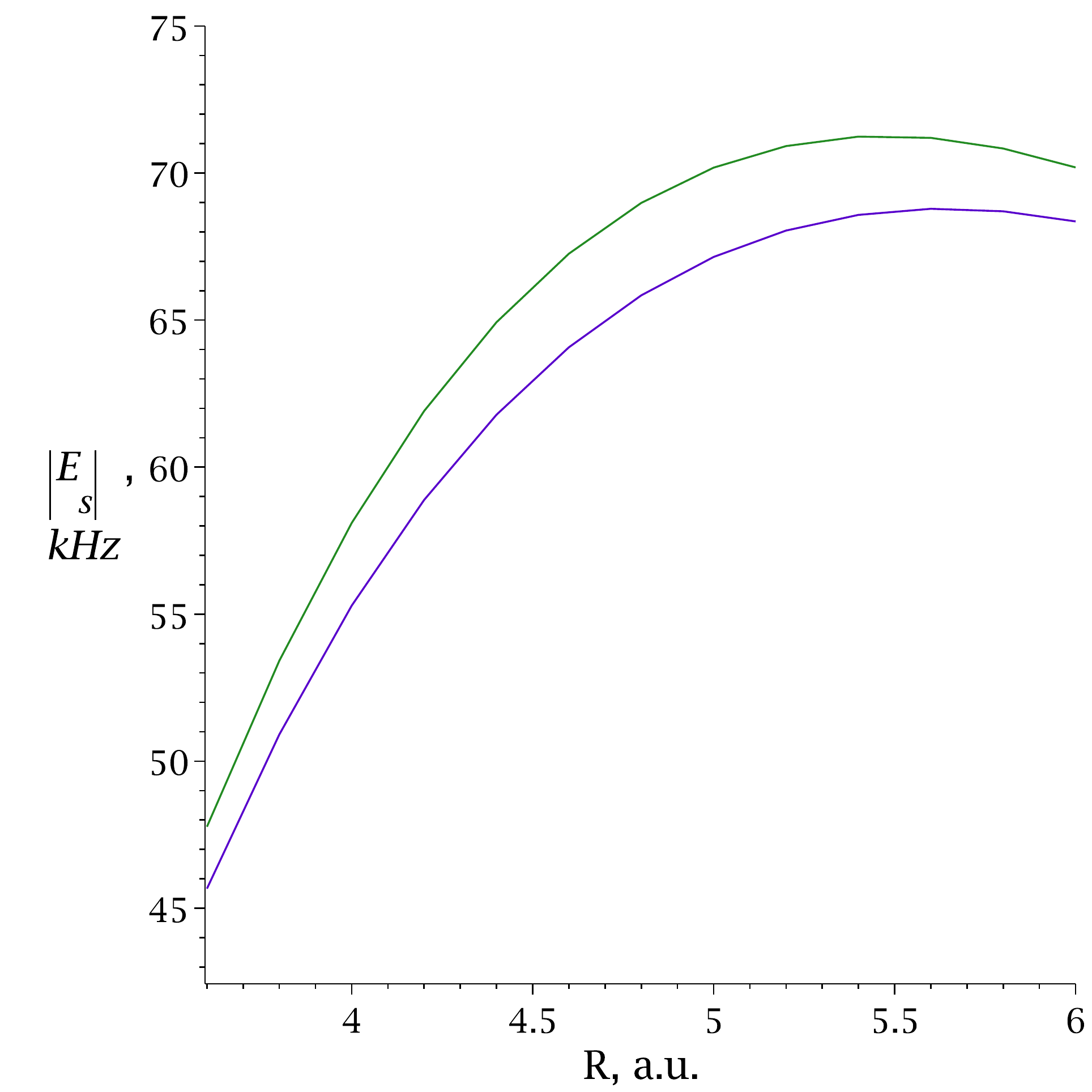}
  \caption{$E_{\rm eff}(R,\theta=0)$ and $E_s(R,\theta=0)$ for RaOH calculated for linear configurations as a function of Ra-OH c.m. distance. Violet line corresponds to SCF values, green line takes into account CCSD correction}
  \label{fgrWdLinear}
\end{figure}

The potential surface computed on a grid for each value of $R$ was interpolated by Akima splines and then expanded in terms of Legendre Polynomials \eqref{LegendreExpansion} with $k_{max}=40$. The coefficients $V_k(R)$ were then interpolated by Akima splines. The bicubic interpolation was applied to the values $E_{\rm eff}$ and $E_s$ computed on a grid. The interpolated functions were used to set up the close-coupled equations for $F_{Jjl}(R)$ by means of a code developed by the authors.

\section{Results and discussion}
The obtained potential surface is represented at Fig. \ref{fgr:Fig.4}. The minimum corresponds to $R=4.4 \,a.u.$ and $\theta=0^\circ$.
The spectrum of the vibrational levels obtained from the solution of the close-coupled equations correspond to the values $\nu_1(\sigma^+)=469 cm^{-1}$ and $\nu_2(\pi)=363 cm^{-1}$ for ${}^2\Sigma_\frac{1}{2}$ ground state. These values can be compared to $\nu_1(\sigma^+)=437 cm^{-1}$ and $\nu_2(\pi)=366 cm^{-1}$ computed in \cite{Isaev_2017} by multi-configurational SCF method.

The difference between energy levels of opposite parities and same quantum numbers $J=1$, $v=1$ gives the value of $l$-doubling equal to $14.467 MHz$. For the energy levels with $J=2$, $v=1$ we obtain the value of $l$-doubling $43.400 MHz$ which is three times larger. This is consistent with theoretical considerations that $l$-doubling is proportional to the factor $J(J+1)$ and confirms the numerical stability of our computation.

The calculated for nonlinear configurations $E_{\rm eff}(R,\theta)$ and $E_s(R,\theta)$ are represented at Fig. \ref{fgrWd} and \ref{fgrWs} respectively and show similar dependence of 
the parameters on the Jacobi coordinates. On Fig. \ref{fgrWdLinear} the results for linear configuration are given.
We compare the values (\ref{Eeffaver}) and (\ref{Esaver}) averaged over the vibrational nuclear wave function  to the values computed for the equilibrium configuration at Table \ref{tbl:tableResults}. It can be seen that the difference from the equilibrium values $E_{\rm eff}$ and $E_s$ is $0.01\%$ for $v=0$ and $0.58\%$ for $v=1$. This means that the results obtained for the equilibrium configuration give good approximation for the lowest vibrational levels. Our result does not confirm the expectations of \cite{prasannaa2019enhanced} based on their YbOH computations that the dependence of the $\mathcal{P}$, $\mathcal{T}$ parameters on the bond angle may significantly affect the observed value.

\begin{table}[h]
\small
  \caption{}
  \label{tbl:tableResults}
  \renewcommand{\arraystretch}{1.5}
  \begin{tabular*}{0.48\textwidth}{@{\extracolsep{\fill}}lll}
    \hline
                    & $E_{\rm eff},\, \frac{\rm GV}{\rm cm}$ &$E_s,\, {\rm kHz}$\\
    \hline
equilibrium  & -48.866  & -64.788\\
$v=0$        & -48.863 & -64.784\\
$v=1$        & -48.585 & -64.416\\
    \hline
cGHF RaOH \cite{gaul2020ab} & -56.87 & -76.5\\
cGKS RaOH \cite{gaul2020ab} & -52.32 & -70.5\\
    \hline
  \end{tabular*}
\end{table}


\section{acknowledgement}
The work is supported by the Russian Science Foundation grant No. 18-12-00227.
We thank to I.P. Kurchavov  for providing equivalent basis sets for Ra atom used for restoration procedure.


%
\bibliographystyle{apsrev}

\begin{thebibliography}{33}
\expandafter\ifx\csname natexlab\endcsname\relax\def\natexlab#1{#1}\fi
\expandafter\ifx\csname bibnamefont\endcsname\relax
  \def\bibnamefont#1{#1}\fi
\expandafter\ifx\csname bibfnamefont\endcsname\relax
  \def\bibfnamefont#1{#1}\fi
\expandafter\ifx\csname citenamefont\endcsname\relax
  \def\citenamefont#1{#1}\fi
\expandafter\ifx\csname url\endcsname\relax
  \def\url#1{\texttt{#1}}\fi
\expandafter\ifx\csname urlprefix\endcsname\relax\def\urlprefix{URL }\fi
\providecommand{\bibinfo}[2]{#2}
\providecommand{\eprint}[2][]{\url{#2}}

\bibitem[{\citenamefont{Khriplovich and Lamoreaux}(2012)}]{khriplovich2012cp}
\bibinfo{author}{\bibfnamefont{I.~B.} \bibnamefont{Khriplovich}}
  \bibnamefont{and} \bibinfo{author}{\bibfnamefont{S.~K.}
  \bibnamefont{Lamoreaux}}, \emph{\bibinfo{title}{CP violation without
  strangeness: electric dipole moments of particles, atoms, and molecules}}
  (\bibinfo{publisher}{Springer Science \& Business Media},
  \bibinfo{year}{2012}).

\bibitem[{\citenamefont{Baron et~al.}(2014)\citenamefont{Baron, Campbell,
  DeMille, Doyle, Gabrielse, Gurevich, Hess, Hutzler, Kirilov, Kozyryev
  et~al.}}]{baron2014order}
\bibinfo{author}{\bibfnamefont{J.}~\bibnamefont{Baron}},
  \bibinfo{author}{\bibfnamefont{W.~C.} \bibnamefont{Campbell}},
  \bibinfo{author}{\bibfnamefont{D.}~\bibnamefont{DeMille}},
  \bibinfo{author}{\bibfnamefont{J.~M.} \bibnamefont{Doyle}},
  \bibinfo{author}{\bibfnamefont{G.}~\bibnamefont{Gabrielse}},
  \bibinfo{author}{\bibfnamefont{Y.~V.} \bibnamefont{Gurevich}},
  \bibinfo{author}{\bibfnamefont{P.~W.} \bibnamefont{Hess}},
  \bibinfo{author}{\bibfnamefont{N.~R.} \bibnamefont{Hutzler}},
  \bibinfo{author}{\bibfnamefont{E.}~\bibnamefont{Kirilov}},
  \bibinfo{author}{\bibfnamefont{I.}~\bibnamefont{Kozyryev}},
  \bibnamefont{et~al.}, \bibinfo{journal}{Science}
  \textbf{\bibinfo{volume}{343}}, \bibinfo{pages}{269} (\bibinfo{year}{2014}).

\bibitem[{\citenamefont{Andreev and Hutzler}(2018)}]{andreev2018improved}
\bibinfo{author}{\bibfnamefont{V.}~\bibnamefont{Andreev}} \bibnamefont{and}
  \bibinfo{author}{\bibfnamefont{N.}~\bibnamefont{Hutzler}},
  \bibinfo{journal}{Nature} \textbf{\bibinfo{volume}{562}},
  \bibinfo{pages}{355} (\bibinfo{year}{2018}).

\bibitem[{\citenamefont{Ginges and Flambaum}(2004)}]{ginges2004violations}
\bibinfo{author}{\bibfnamefont{J.}~\bibnamefont{Ginges}} \bibnamefont{and}
  \bibinfo{author}{\bibfnamefont{V.~V.} \bibnamefont{Flambaum}},
  \bibinfo{journal}{Physics Reports} \textbf{\bibinfo{volume}{397}},
  \bibinfo{pages}{63} (\bibinfo{year}{2004}).

\bibitem[{\citenamefont{Maison et~al.}(2019)\citenamefont{Maison, Skripnikov,
  and Flambaum}}]{maison2019theoretical}
\bibinfo{author}{\bibfnamefont{D.~E.} \bibnamefont{Maison}},
  \bibinfo{author}{\bibfnamefont{L.~V.} \bibnamefont{Skripnikov}},
  \bibnamefont{and} \bibinfo{author}{\bibfnamefont{V.~V.}
  \bibnamefont{Flambaum}}, \bibinfo{journal}{Physical Review A}
  \textbf{\bibinfo{volume}{100}}, \bibinfo{pages}{032514}
  (\bibinfo{year}{2019}).

\bibitem[{\citenamefont{Maison et~al.}(2020)\citenamefont{Maison, Flambaum,
  Hutzler, and Skripnikov}}]{maison2020study}
\bibinfo{author}{\bibfnamefont{D.}~\bibnamefont{Maison}},
  \bibinfo{author}{\bibfnamefont{V.}~\bibnamefont{Flambaum}},
  \bibinfo{author}{\bibfnamefont{N.}~\bibnamefont{Hutzler}}, \bibnamefont{and}
  \bibinfo{author}{\bibfnamefont{L.}~\bibnamefont{Skripnikov}},
  \bibinfo{journal}{arXiv preprint arXiv:2010.11669}  (\bibinfo{year}{2020}).

\bibitem[{\citenamefont{Isaev et~al.}(2010)\citenamefont{Isaev, Hoekstra, and
  Berger}}]{Isaev:2010}
\bibinfo{author}{\bibfnamefont{T.~A.} \bibnamefont{Isaev}},
  \bibinfo{author}{\bibfnamefont{S.}~\bibnamefont{Hoekstra}}, \bibnamefont{and}
  \bibinfo{author}{\bibfnamefont{R.}~\bibnamefont{Berger}},
  \bibinfo{journal}{Phys.\ Rev.\ A} \textbf{\bibinfo{volume}{82}},
  \bibinfo{pages}{052521} (\bibinfo{year}{2010}).

\bibitem[{\citenamefont{Garcia~Ruiz et~al.}(2020)\citenamefont{Garcia~Ruiz,
  Berger, Billowes, Binnersley, Bissell, Breier, Brinson, Chrysalidis,
  Cocolios, Cooper et~al.}}]{GarciaRuiz2020}
\bibinfo{author}{\bibfnamefont{R.~F.} \bibnamefont{Garcia~Ruiz}},
  \bibinfo{author}{\bibfnamefont{R.}~\bibnamefont{Berger}},
  \bibinfo{author}{\bibfnamefont{J.}~\bibnamefont{Billowes}},
  \bibinfo{author}{\bibfnamefont{C.~L.} \bibnamefont{Binnersley}},
  \bibinfo{author}{\bibfnamefont{M.~L.} \bibnamefont{Bissell}},
  \bibinfo{author}{\bibfnamefont{A.~A.} \bibnamefont{Breier}},
  \bibinfo{author}{\bibfnamefont{A.~J.} \bibnamefont{Brinson}},
  \bibinfo{author}{\bibfnamefont{K.}~\bibnamefont{Chrysalidis}},
  \bibinfo{author}{\bibfnamefont{T.~E.} \bibnamefont{Cocolios}},
  \bibinfo{author}{\bibfnamefont{B.~S.} \bibnamefont{Cooper}},
  \bibnamefont{et~al.}, \bibinfo{journal}{Nature}
  \textbf{\bibinfo{volume}{581}}, \bibinfo{pages}{396} (\bibinfo{year}{2020}),
  ISSN \bibinfo{issn}{1476-4687},
  \urlprefix\url{https://doi.org/10.1038/s41586-020-2299-4}.

\bibitem[{\citenamefont{Lim et~al.}(2018)\citenamefont{Lim, Almond, Trigatzis,
  Devlin, Fitch, Sauer, Tarbutt, and Hinds}}]{YbFLaserCooled}
\bibinfo{author}{\bibfnamefont{J.}~\bibnamefont{Lim}},
  \bibinfo{author}{\bibfnamefont{J.~R.} \bibnamefont{Almond}},
  \bibinfo{author}{\bibfnamefont{M.~A.} \bibnamefont{Trigatzis}},
  \bibinfo{author}{\bibfnamefont{J.~A.} \bibnamefont{Devlin}},
  \bibinfo{author}{\bibfnamefont{N.~J.} \bibnamefont{Fitch}},
  \bibinfo{author}{\bibfnamefont{B.~E.} \bibnamefont{Sauer}},
  \bibinfo{author}{\bibfnamefont{M.~R.} \bibnamefont{Tarbutt}},
  \bibnamefont{and} \bibinfo{author}{\bibfnamefont{E.~A.} \bibnamefont{Hinds}},
  \bibinfo{journal}{Phys. Rev. Lett.} \textbf{\bibinfo{volume}{120}},
  \bibinfo{pages}{123201} (\bibinfo{year}{2018}),
  \urlprefix\url{https://link.aps.org/doi/10.1103/PhysRevLett.120.123201}.

\bibitem[{\citenamefont{Andreev et~al.}(2018)\citenamefont{Andreev, Ang,
  DeMille, Doyle, Gabrielse, Haefner, Hutzler, Lasner, Meisenhelder, O'Leary
  et~al.}}]{ACME:18}
\bibinfo{author}{\bibfnamefont{V.}~\bibnamefont{Andreev}},
  \bibinfo{author}{\bibfnamefont{D.}~\bibnamefont{Ang}},
  \bibinfo{author}{\bibfnamefont{D.}~\bibnamefont{DeMille}},
  \bibinfo{author}{\bibfnamefont{J.}~\bibnamefont{Doyle}},
  \bibinfo{author}{\bibfnamefont{G.}~\bibnamefont{Gabrielse}},
  \bibinfo{author}{\bibfnamefont{J.}~\bibnamefont{Haefner}},
  \bibinfo{author}{\bibfnamefont{N.}~\bibnamefont{Hutzler}},
  \bibinfo{author}{\bibfnamefont{Z.}~\bibnamefont{Lasner}},
  \bibinfo{author}{\bibfnamefont{C.}~\bibnamefont{Meisenhelder}},
  \bibinfo{author}{\bibfnamefont{B.}~\bibnamefont{O'Leary}},
  \bibnamefont{et~al.}, \bibinfo{journal}{Nature}
  \textbf{\bibinfo{volume}{562}}, \bibinfo{pages}{355} (\bibinfo{year}{2018}).

\bibitem[{\citenamefont{DeMille et~al.}(2001)\citenamefont{DeMille,
  an~S.~Bickman, Kawall, Hunter, Krause, Jr, Maxwell, and
  Ulmer}}]{DeMille:2001}
\bibinfo{author}{\bibfnamefont{D.}~\bibnamefont{DeMille}},
  \bibinfo{author}{\bibfnamefont{F.~B.} \bibnamefont{an~S.~Bickman}},
  \bibinfo{author}{\bibfnamefont{D.}~\bibnamefont{Kawall}},
  \bibinfo{author}{\bibfnamefont{L.}~\bibnamefont{Hunter}},
  \bibinfo{author}{\bibfnamefont{D.}~\bibnamefont{Krause}},
  \bibinfo{author}{\bibnamefont{Jr}},
  \bibinfo{author}{\bibfnamefont{S.}~\bibnamefont{Maxwell}}, \bibnamefont{and}
  \bibinfo{author}{\bibfnamefont{K.}~\bibnamefont{Ulmer}},
  \bibinfo{journal}{AIP Conf. Proc.} \textbf{\bibinfo{volume}{596}},
  \bibinfo{pages}{72} (\bibinfo{year}{2001}).

\bibitem[{\citenamefont{Petrov et~al.}(2014)\citenamefont{Petrov, Skripnikov,
  Titov, Hutzler, Hess, O'Leary, Spaun, DeMille, Gabrielse, and
  Doyle}}]{Petrov:14}
\bibinfo{author}{\bibfnamefont{A.~N.} \bibnamefont{Petrov}},
  \bibinfo{author}{\bibfnamefont{L.~V.} \bibnamefont{Skripnikov}},
  \bibinfo{author}{\bibfnamefont{A.~V.} \bibnamefont{Titov}},
  \bibinfo{author}{\bibfnamefont{N.~R.} \bibnamefont{Hutzler}},
  \bibinfo{author}{\bibfnamefont{P.~W.} \bibnamefont{Hess}},
  \bibinfo{author}{\bibfnamefont{B.~R.} \bibnamefont{O'Leary}},
  \bibinfo{author}{\bibfnamefont{B.}~\bibnamefont{Spaun}},
  \bibinfo{author}{\bibfnamefont{D.}~\bibnamefont{DeMille}},
  \bibinfo{author}{\bibfnamefont{G.}~\bibnamefont{Gabrielse}},
  \bibnamefont{and} \bibinfo{author}{\bibfnamefont{J.~M.} \bibnamefont{Doyle}},
  \bibinfo{journal}{Phys. Rev. A} \textbf{\bibinfo{volume}{89}},
  \bibinfo{pages}{062505} (\bibinfo{year}{2014}).

\bibitem[{\citenamefont{Vutha et~al.}(2010)\citenamefont{Vutha, Campbell,
  Gurevich, Hutzler, Parsons, Patterson, Petrik, Spaun, Doyle, Gabrielse
  et~al.}}]{Vutha:2010}
\bibinfo{author}{\bibfnamefont{A.~C.} \bibnamefont{Vutha}},
  \bibinfo{author}{\bibfnamefont{W.~C.} \bibnamefont{Campbell}},
  \bibinfo{author}{\bibfnamefont{Y.~V.} \bibnamefont{Gurevich}},
  \bibinfo{author}{\bibfnamefont{N.~R.} \bibnamefont{Hutzler}},
  \bibinfo{author}{\bibfnamefont{M.}~\bibnamefont{Parsons}},
  \bibinfo{author}{\bibfnamefont{D.}~\bibnamefont{Patterson}},
  \bibinfo{author}{\bibfnamefont{E.}~\bibnamefont{Petrik}},
  \bibinfo{author}{\bibfnamefont{B.}~\bibnamefont{Spaun}},
  \bibinfo{author}{\bibfnamefont{J.~M.} \bibnamefont{Doyle}},
  \bibinfo{author}{\bibfnamefont{G.}~\bibnamefont{Gabrielse}},
  \bibnamefont{et~al.}, \bibinfo{journal}{J.\ Phys.\ B}
  \textbf{\bibinfo{volume}{43}}, \bibinfo{pages}{074007}
  (\bibinfo{year}{2010}).

\bibitem[{\citenamefont{Petrov}(2015)}]{Petrov:15}
\bibinfo{author}{\bibfnamefont{A.~N.} \bibnamefont{Petrov}},
  \bibinfo{journal}{Phys.\ Rev.\ A} \textbf{\bibinfo{volume}{91}},
  \bibinfo{pages}{062509} (\bibinfo{year}{2015}).

\bibitem[{\citenamefont{Petrov}(2017)}]{Petrov:17}
\bibinfo{author}{\bibfnamefont{A.~N.} \bibnamefont{Petrov}},
  \bibinfo{journal}{Phys. Rev. A} \textbf{\bibinfo{volume}{95}},
  \bibinfo{pages}{062501} (\bibinfo{year}{2017}),
  \urlprefix\url{https://link.aps.org/doi/10.1103/PhysRevA.95.062501}.

\bibitem[{\citenamefont{Cairncross et~al.}(2017)\citenamefont{Cairncross,
  Gresh, Grau, Cossel, Roussy, Ni, Zhou, Ye, and Cornell}}]{Cornell:2017}
\bibinfo{author}{\bibfnamefont{W.~B.} \bibnamefont{Cairncross}},
  \bibinfo{author}{\bibfnamefont{D.~N.} \bibnamefont{Gresh}},
  \bibinfo{author}{\bibfnamefont{M.}~\bibnamefont{Grau}},
  \bibinfo{author}{\bibfnamefont{K.~C.} \bibnamefont{Cossel}},
  \bibinfo{author}{\bibfnamefont{T.~S.} \bibnamefont{Roussy}},
  \bibinfo{author}{\bibfnamefont{Y.}~\bibnamefont{Ni}},
  \bibinfo{author}{\bibfnamefont{Y.}~\bibnamefont{Zhou}},
  \bibinfo{author}{\bibfnamefont{J.}~\bibnamefont{Ye}}, \bibnamefont{and}
  \bibinfo{author}{\bibfnamefont{E.~A.} \bibnamefont{Cornell}},
  \bibinfo{journal}{Phys.\ Rev.\ Lett.} \textbf{\bibinfo{volume}{119}},
  \bibinfo{pages}{153001} (\bibinfo{year}{2017}).

\bibitem[{\citenamefont{Petrov}(2018)}]{Petrov:18}
\bibinfo{author}{\bibfnamefont{A.~N.} \bibnamefont{Petrov}},
  \bibinfo{journal}{Phys. Rev. A} \textbf{\bibinfo{volume}{97}},
  \bibinfo{pages}{052504} (\bibinfo{year}{2018}).

\bibitem[{\citenamefont{Isaev et~al.}(2017)\citenamefont{Isaev, Zaitsevskii,
  and Eliav}}]{Isaev_2017}
\bibinfo{author}{\bibfnamefont{T.~A.} \bibnamefont{Isaev}},
  \bibinfo{author}{\bibfnamefont{A.~V.} \bibnamefont{Zaitsevskii}},
  \bibnamefont{and} \bibinfo{author}{\bibfnamefont{E.}~\bibnamefont{Eliav}},
  \bibinfo{journal}{Journal of Physics B: Atomic, Molecular and Optical
  Physics} \textbf{\bibinfo{volume}{50}}, \bibinfo{pages}{225101}
  (\bibinfo{year}{2017}),
  \urlprefix\url{https://doi.org/10.1088%2F1361-6455%2Faa8f34}.

\bibitem[{\citenamefont{Kozyryev and Hutzler}(2017)}]{Kozyryev:17}
\bibinfo{author}{\bibfnamefont{I.}~\bibnamefont{Kozyryev}} \bibnamefont{and}
  \bibinfo{author}{\bibfnamefont{N.~R.} \bibnamefont{Hutzler}},
  \bibinfo{journal}{Phys. Rev. Lett.} \textbf{\bibinfo{volume}{119}},
  \bibinfo{pages}{133002} (\bibinfo{year}{2017}),
  \urlprefix\url{https://link.aps.org/doi/10.1103/PhysRevLett.119.133002}.

\bibitem[{\citenamefont{Hutzler}(2020)}]{hutzler2020polyatomic}
\bibinfo{author}{\bibfnamefont{N.~R.} \bibnamefont{Hutzler}},
  \bibinfo{journal}{arXiv preprint arXiv:2008.03398}  (\bibinfo{year}{2020}).

\bibitem[{\citenamefont{Denis et~al.}(2019)\citenamefont{Denis, Haase,
  Timmermans, Eliav, Hutzler, and Borschevsky}}]{denis2019enhancement}
\bibinfo{author}{\bibfnamefont{M.}~\bibnamefont{Denis}},
  \bibinfo{author}{\bibfnamefont{P.~A.} \bibnamefont{Haase}},
  \bibinfo{author}{\bibfnamefont{R.~G.} \bibnamefont{Timmermans}},
  \bibinfo{author}{\bibfnamefont{E.}~\bibnamefont{Eliav}},
  \bibinfo{author}{\bibfnamefont{N.~R.} \bibnamefont{Hutzler}},
  \bibnamefont{and}
  \bibinfo{author}{\bibfnamefont{A.}~\bibnamefont{Borschevsky}},
  \bibinfo{journal}{Physical Review A} \textbf{\bibinfo{volume}{99}},
  \bibinfo{pages}{042512} (\bibinfo{year}{2019}).

\bibitem[{\citenamefont{Prasannaa et~al.}(2019)\citenamefont{Prasannaa,
  Shitara, Sakurai, Abe, and Das}}]{prasannaa2019enhanced}
\bibinfo{author}{\bibfnamefont{V.}~\bibnamefont{Prasannaa}},
  \bibinfo{author}{\bibfnamefont{N.}~\bibnamefont{Shitara}},
  \bibinfo{author}{\bibfnamefont{A.}~\bibnamefont{Sakurai}},
  \bibinfo{author}{\bibfnamefont{M.}~\bibnamefont{Abe}}, \bibnamefont{and}
  \bibinfo{author}{\bibfnamefont{B.}~\bibnamefont{Das}},
  \bibinfo{journal}{Physical Review A} \textbf{\bibinfo{volume}{99}},
  \bibinfo{pages}{062502} (\bibinfo{year}{2019}).

\bibitem[{\citenamefont{Gaul and Berger}(2020)}]{gaul2020ab}
\bibinfo{author}{\bibfnamefont{K.}~\bibnamefont{Gaul}} \bibnamefont{and}
  \bibinfo{author}{\bibfnamefont{R.}~\bibnamefont{Berger}},
  \bibinfo{journal}{Physical Review A} \textbf{\bibinfo{volume}{101}},
  \bibinfo{pages}{012508} (\bibinfo{year}{2020}).

\bibitem[{\citenamefont{McGuire and Kouri}(1974)}]{mcguire1974quantum}
\bibinfo{author}{\bibfnamefont{P.}~\bibnamefont{McGuire}} \bibnamefont{and}
  \bibinfo{author}{\bibfnamefont{D.~J.} \bibnamefont{Kouri}},
  \bibinfo{journal}{The Journal of Chemical Physics}
  \textbf{\bibinfo{volume}{60}}, \bibinfo{pages}{2488} (\bibinfo{year}{1974}).

\bibitem[{DIR()}]{DIRAC19}
\bibinfo{note}{{DIRAC}, a relativistic ab initio electronic structure program,
  Release {DIRAC19} (2019), written by A.~S.~P.~Gomes, T.~Saue, L.~Visscher,
  H.~J.~{\relax Aa}.~Jensen, and R.~Bast, with contributions from I.~A.~Aucar,
  V.~Bakken, K.~G.~Dyall, S.~Dubillard, U.~Ekstr{\"o}m, E.~Eliav,
  T.~Enevoldsen, E.~Fa{\ss}hauer, T.~Fleig, O.~Fossgaard, L.~Halbert,
  E.~D.~Hedeg{\aa}rd, B.~Heimlich--Paris, T.~Helgaker, J.~Henriksson,
  M.~Ilia{\v{s}}, Ch.~R.~Jacob, S.~Knecht, S.~Komorovsk{\'y}, O.~Kullie,
  J.~K.~L{\ae}rdahl, C.~V.~Larsen, Y.~S.~Lee, H.~S.~Nataraj, M.~K.~Nayak,
  P.~Norman, G.~Olejniczak, J.~Olsen, J.~M.~H.~Olsen, Y.~C.~Park,
  J.~K.~Pedersen, M.~Pernpointner, R.~di~Remigio, K.~Ruud, P.~Sa{\l}ek,
  B.~Schimmelpfennig, B.~Senjean, A.~Shee, J.~Sikkema, A.~J.~Thorvaldsen,
  J.~Thyssen, J.~van~Stralen, M.~L.~Vidal, S.~Villaume, O.~Visser, T.~Winther,
  and S.~Yamamoto (available at http://dx.doi.org/10.5281/zenodo.3572669, see
  also http://www.diracprogram.org)}.

\bibitem[{\citenamefont{Titov and Mosyagin}(1999)}]{titov1999generalized}
\bibinfo{author}{\bibfnamefont{A.}~\bibnamefont{Titov}} \bibnamefont{and}
  \bibinfo{author}{\bibfnamefont{N.}~\bibnamefont{Mosyagin}},
  \bibinfo{journal}{International journal of quantum chemistry}
  \textbf{\bibinfo{volume}{71}}, \bibinfo{pages}{359} (\bibinfo{year}{1999}).

\bibitem[{\citenamefont{Mosyagin et~al.}(2010)\citenamefont{Mosyagin,
  Zaitsevskii, and Titov}}]{mosyagin2010shape}
\bibinfo{author}{\bibfnamefont{N.~S.} \bibnamefont{Mosyagin}},
  \bibinfo{author}{\bibfnamefont{A.}~\bibnamefont{Zaitsevskii}},
  \bibnamefont{and} \bibinfo{author}{\bibfnamefont{A.~V.} \bibnamefont{Titov}},
  \bibinfo{journal}{International Review of Atomic and Molecular Physics}
  \textbf{\bibinfo{volume}{1}}, \bibinfo{pages}{63} (\bibinfo{year}{2010}).

\bibitem[{\citenamefont{Mosyagin et~al.}(2016)\citenamefont{Mosyagin,
  Zaitsevskii, Skripnikov, and Titov}}]{mosyagin2016generalized}
\bibinfo{author}{\bibfnamefont{N.~S.} \bibnamefont{Mosyagin}},
  \bibinfo{author}{\bibfnamefont{A.~V.} \bibnamefont{Zaitsevskii}},
  \bibinfo{author}{\bibfnamefont{L.~V.} \bibnamefont{Skripnikov}},
  \bibnamefont{and} \bibinfo{author}{\bibfnamefont{A.~V.} \bibnamefont{Titov}},
  \bibinfo{journal}{International Journal of Quantum Chemistry}
  \textbf{\bibinfo{volume}{116}}, \bibinfo{pages}{301} (\bibinfo{year}{2016}).

\bibitem[{\citenamefont{{URL:
  http://www.qchem.pnpi.spb.ru/Basis/~}}()}]{QCPNPI:Basis}
\bibinfo{author}{\bibnamefont{{URL: http://www.qchem.pnpi.spb.ru/Basis/~}}},
  \bibinfo{note}{~{GRECPs} and basis sets}.

\bibitem[{\citenamefont{Petrov et~al.}(2002)\citenamefont{Petrov, Mosyagin,
  Isaev, Titov, Ezhov, Eliav, and Kaldor}}]{Petrov:02}
\bibinfo{author}{\bibfnamefont{A.~N.} \bibnamefont{Petrov}},
  \bibinfo{author}{\bibfnamefont{N.~S.} \bibnamefont{Mosyagin}},
  \bibinfo{author}{\bibfnamefont{T.~A.} \bibnamefont{Isaev}},
  \bibinfo{author}{\bibfnamefont{A.~V.} \bibnamefont{Titov}},
  \bibinfo{author}{\bibfnamefont{V.~F.} \bibnamefont{Ezhov}},
  \bibinfo{author}{\bibfnamefont{E.}~\bibnamefont{Eliav}}, \bibnamefont{and}
  \bibinfo{author}{\bibfnamefont{U.}~\bibnamefont{Kaldor}},
  \bibinfo{journal}{Phys.\ Rev.\ Lett.} \textbf{\bibinfo{volume}{88}},
  \bibinfo{pages}{073001} (\bibinfo{year}{2002}).

\bibitem[{\citenamefont{Titov et~al.}(2006)\citenamefont{Titov, Mosyagin,
  Petrov, and Isaev}}]{titov2006d}
\bibinfo{author}{\bibfnamefont{A.}~\bibnamefont{Titov}},
  \bibinfo{author}{\bibfnamefont{N.}~\bibnamefont{Mosyagin}},
  \bibinfo{author}{\bibfnamefont{A.}~\bibnamefont{Petrov}}, \bibnamefont{and}
  \bibinfo{author}{\bibfnamefont{T.}~\bibnamefont{Isaev}},
  \bibinfo{journal}{Chem. Phys} \textbf{\bibinfo{volume}{15}},
  \bibinfo{pages}{253} (\bibinfo{year}{2006}).

\bibitem[{\citenamefont{Skripnikov and
  Titov}(2015)}]{skripnikov2015theoretical}
\bibinfo{author}{\bibfnamefont{L.}~\bibnamefont{Skripnikov}} \bibnamefont{and}
  \bibinfo{author}{\bibfnamefont{A.}~\bibnamefont{Titov}},
  \bibinfo{journal}{Physical Review A} \textbf{\bibinfo{volume}{91}},
  \bibinfo{pages}{042504} (\bibinfo{year}{2015}).

\bibitem[{MRC()}]{MRCC2020}
\bibinfo{note}{M. K\'{a}llay, P. R. Nagy, D. Mester, Z. Rolik, G. Samu, J.
  Csontos, J. Cs\'{o}ka, P. B. Szab\'{o}, L. Gyevi-Nagy, B. H\'{e}gely, I.
  Ladj\'{a}nszki, L. Szegedy, B. Lad\'{o}czki, K. Petrov, M. Farkas, P. D.
  Mezei, and \'{a}. Ganyecz: The {\sc mrcc} program system: Accurate quantum
  chemistry from water to proteins, J. Chem. Phys. 152, 074107 (2020).
  {\sc mrcc}, a quantum chemical program suite written by M. K\'{a}llay, P.
  R. Nagy, D. Mester, Z. Rolik, G. Samu, J. Csontos, J. Cs\'{o}ka, P. B.
  Szab\'{o}, L. Gyevi-Nagy, B. H\'{e}gely, I. Ladj\'{a}nszki, L. Szegedy, B.
  Lad\'{o}czki, K. Petrov, M. Farkas, P. D. Mezei, and \'{a}. Ganyecz. See www.mrcc.hu.
  }
\end{thebibliography}
%


\end{document}